%% file: arxiv.tex
\documentclass[10pt,twocolumn]{article} 
\usepackage{simpleConference}
\usepackage{times}
\usepackage{graphicx}
\usepackage{amssymb}
\usepackage{url,hyperref}
\usepackage{authblk}
\usepackage{amsmath,amssymb,amsfonts, bm}
\usepackage{subfigure}
\begin{document}

\title{Crowd simulation incorporating a route choice model \\ and similarity evaluation using real large-scale data}

\author[1, 2, *]{Ryo Nishida}
\author[2]{Masaki Onishi}
\author[1]{Koichi Hashimoto}
\affil[1]{Tohoku University}
\affil[2]{National Institute of Advanced Industrial Science and Technology}
\affil[*]{\text{ryo.nishida.t4@dc.tohoku.ac.jp}}

\maketitle
\thispagestyle{empty}

\begin{abstract}
Modeling and simulation approaches that express crowd movement with mathematical models are widely and actively studied to understand crowd movement and resolve crowd accidents. Existing literature on crowd modeling focuses on only the decision-making of walking behavior. However, the decision-making of route choice, which is a higher-level decision, should also be modeled for constructing more practical simulations. Furthermore, the reproducibility evaluation of the crowd simulation incorporating the route choice model using real data is insufficient. Therefore, we generalize and propose a crowd simulation framework that includes actual crowd movement measurements, route choice model estimation, and crowd simulator construction. We use the Discrete choice model as the route choice model and the Social force model as the walking model. In experiments, we measure crowd movements during an evacuation drill in a theater and a firework event where tens of thousands of people moved and prove that the crowd simulation incorporating the route choice model can reproduce the real large-scale crowd movement more accurately.
\end{abstract}


\section{Introduction}
Crowd movement is an emergent phenomenon that results from the interaction of pedestrians. Many people gather in the same place and move individually, simultaneously, during disasters such as earthquakes and fires and large-scale events such as music concerts and fireworks events. Stampedes may occur if someone collapses during crowd turbulence, leading to a severe crowd accident. So far, large-scale crowd movements have sparked crowd accidents \cite{Soomaroo2012,Keith}.

We need to understand how pedestrians move and what crowd behavior has emerged to resolve crowd accidents. To understand crowd movement, modeling and simulation approaches, expressing the crowd movement with a mathematical model, are widely and actively studied. Here, we briefly explain the modeling approach. For more details of these models, we refer the readers to \cite{Yang2020,Zhou2018,Sidiropoulos2020,Duives2013}. 

\begin{figure}[tb]
\centering
\includegraphics[width=1.0\linewidth]{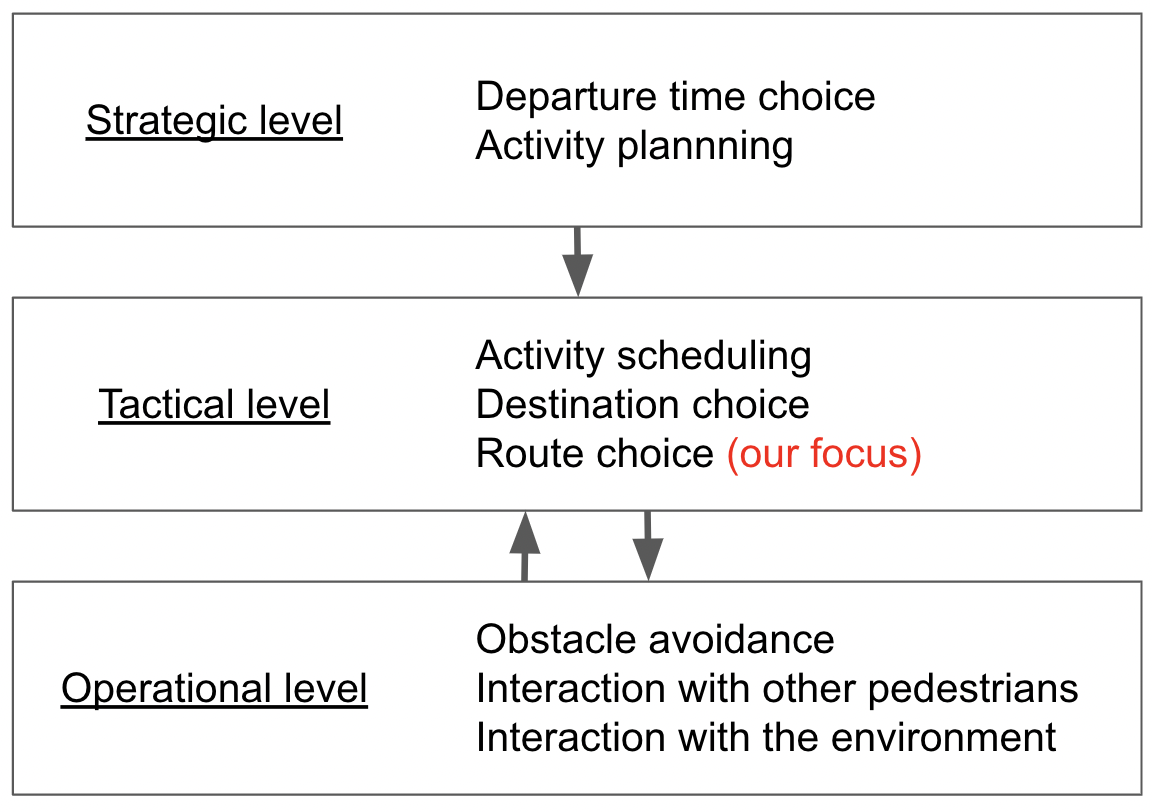}
\caption{Pedestrian behavior level by \cite{Hoogendoorn2004}}
\label{fig:level}
\end{figure}

The models used in crowd simulation can be classified into macroscopic and microscopic models. 
Macroscopic models do not focus on the movements of individual pedestrians but treat crowds in a unified and continuous manner. The crowd movement is reproduced employing the knowledge of fluid dynamics and potential fields \cite{Lu2016-vb,Narain2009-pa}. On the other hand, microscopic models focus on the movement of individual pedestrians and represents the interactions between pedestrians and the environment e.g., collision avoidance. 
Since 1990s, many researchers have proposed microscopic models such as the Cellular automata model (CA) \cite{Blue1999,Fukui1999}, Social force model (SFM) \cite{Helbing1995,Helbing2000b}, and Reciprocal velocity obstacle (RVO) \cite{rvo}.
CA divides the space into lattice cells and considers the movement of pedestrians as the movement between the cells. 
SFM is a physical model that treats pedestrians as a point mass. The forces received from other pedestrians and the environment are defined as Social Force, and the movement of pedestrians is formulated using these forces. 
RVO is an extension of Velocity obstacles (VO) to crowd simulation. VO is an algorithm in robotics that determines the velocity of a moving object so that it does not collide with other moving objects.
These microscopic models are also used for research in the autonomous agents and multiagent systems field \cite{Kullu2017-nn,Crociani2018-vg,Shigenaka2019-qi,Predhumeau2021-as}.

Although earlier studies have made it possible to simulate crowd movement, challenges still remain.
Existing literature focus on only an {\it operational level}, in the pedestrian behavior levels defined by Hoogendoorn et al. \cite{Hoogendoorn2004}. They define pedestrian behavior at three behavior levels, as shown in Figure~\ref{fig:level}. Behaviors are decided, in order, from the top to bottom. Microscopic models such as CA, SFM, and RVO are {\it operational level} models that correspondent to the {\it operational level}, and in almost simulation studies, the destination or route is pre-fixed. However, pedestrians, in general, change their routes depending on the situation. Therefore, the decision-making of route choice should also be modeled in order to perform a more practical simulation.

Modeling and simulation of route choice corresponding to the {\it tactical level} has increased since 2010.
However, the estimation of route choice models and the reproducibility evaluation of the crowd simulation incorporating the route choice model using real data are insufficient.
Organizing modeling and simulation methodologies using real data and evaluating the reproducibility of simulations with real data are necessary for further refinement of crowd simulations.

Therefore, this study aims to propose and evaluate a crowd simulation incorporating a route choice model based on real crowd movement data. Our contributions are: 1) generalizing and proposing a crowd simulation framework that includes actual crowd movement measurements, route choice model estimation, and crowd simulator construction, and 2) measuring the crowd movements during an evacuation drill in a theater and a firework event in which tens of thousands of people moved and verifying the reproducibility of the crowd simulation incorporating the route choice model using the measured real large-scale crowd movement data.

The remainder of this paper is organized as follows. Section 2 reviews studies that incorporate route choice models into crowd simulation. Section 3 describes our crowd simulations that incorporate a route choice model. Section 4 mentions the reproduction of crowd movement during evacuation in a theater, and Section 5 describes the reproduction of crowd movement at a large-scale fireworks event. Finally, Section 6 provides a summary and discusses future work.

\section{Related works}


In this section, we refer to several studies on route choice models in crowd simulation. 
The simplest model is to select the shortest path \cite{Asano2010,Wagoum2012}. However, the shortest path selection model (SP) cannot completely reproduce the actual crowd movement. Because pedestrians take into account factors such as route attractiveness, route width, congestion, and habits, in addition to distance \cite{Agrawal2008-za,Guo2013-wc,Lovreglio2016-yg}. Therefore, utility-based modeling that unifies the combination of factors involved in route choice in the form of utility (or cost) and assumes that the route with the maximum utility is chosen, is widely used. Several models have been proposed that define their own utility functions and selection methods \cite{Stubenschrott2014,Crociani2016,Liao2017,Vizzari2020-zs}. In general, utility-based models have been formulated and widely used as Discrete choice model (DCM). The simulation of crowd movement, including route choice, is being proposed in combination with DCM as a route choice model and SFM or system dynamics model as a walking model \cite{Liu2012-aw,Lovreglio2016-yg,Haghani2017-bv,Haghani2019-el,Gao2019-mx,Yang2020-uj,Lu2021-fk}.



First, there are some studies in which the researchers determined the DCM parameters and combined DCM with crowd simulators.
For example, Liu et al. modeled the route choice using a DCM with distance to the exit, attractiveness to the store, and movement of other people as factors \cite{Liu2012-aw}. SFM and RVO were used as the walking model. Crowd simulations are performed for evacuation, shopping, and movement during a riot. They reported that crowd simulation with DCM represents more complex crowd movement than with SP or random selection. Yang et al. simulated the crowd moving with the guide (navigator) during an evacuation \cite{Yang2020-uj}.
SFM was used as the walking model. The destination of SFM was the location of the guide, and the part that selects which guide to follow was represented by DCM. The simulation was performed by changing the initial position of the guide, and the optimal initial position of the guide for efficient evacuation was obtained. Their study suggested the effectiveness of introducing a route choice model and provided examples of advanced simulation applications. 
However, the parameters of DCM were given by the authors, and the simulation results were not evaluated using real data, so it is unclear whether the model and simulation reproduce realistic crowd movements.

In addition, some studies collect route choice behavior through questionnaire surveys and surveys using virtual reality (VR), estimate DCM parameters, and incorporate them into simulations.
Lovreglio et al. collected data on exit choice through questionnaires and estimate DCM \cite{Lovreglio2016-yg}. The elements of DCM include the distance to the exit, the number of people near the exit, and the number of people in the vicinity of the decision-maker. It is reported that the simulation can represent exit choice bias.
Lu et al. collected data on route choice behavior in a VR experiment and estimated DCM \cite{Lu2021-fk}. The participants in the experiment wore head-mounted displays to virtually experience an evacuation. The factors are distance to the exit, density, and route guidance. They report that finding more effective guidance methods can be accomplished using crowd simulation with response to guidance. 
Data collection through VR experiments is an effective means, but there are concerns that the data may differ from actual behavior.
These studies also did not evaluate the reproducibility of simulations.

Other methods have been proposed to estimate the DCM parameters so that the simulated crowd movement matches the actual crowd movement.
Gao et al. simulated the choice behavior of ticket gates at train stations and compared it with actual measured data. As in previous studies, the walking model was SFM and the route choice model was DCM. However, the parameters of DCM were adjusted to match the simulation results to reality and were not estimated using actual choice behavior data. In this case, it is possible that errors in the simulator were rounded into the parameters of the route choice model, and the model did not represent pure route choice behavior.

The work of Haghani et al. is closest to our goal. They conducted a subject experiment at a sports center, collected data on choice behavior, and used them to estimate the parameters of DCM. The experiments imitated emergency exit route decision-making when escaping a threat in rooms with multiple exit alternatives \cite{Haghani2017-bv}. 
The estimation results of DCM suggest that the movements of large numbers of pedestrians to the route increase the choice probability of the route when the route or environment is partially unknown to the decision-maker.
They also reported that the crowd simulation in which the estimated DCM and SFM combined can represent the crowd movement as in the experiment, from the perspective of the evacuation time \cite{Haghani2019-el}. 
Modeling route (or exit) choice with actual crowd movement data can provide a better understanding of crowd movement. The modeling with data obtained in more realistic situations may solidify the understanding, and the evaluation using real crowd movement data ensures the effectiveness of the crowd simulations. 

\begin{figure*}[tb]
\centering
\includegraphics[width=1.0\linewidth]{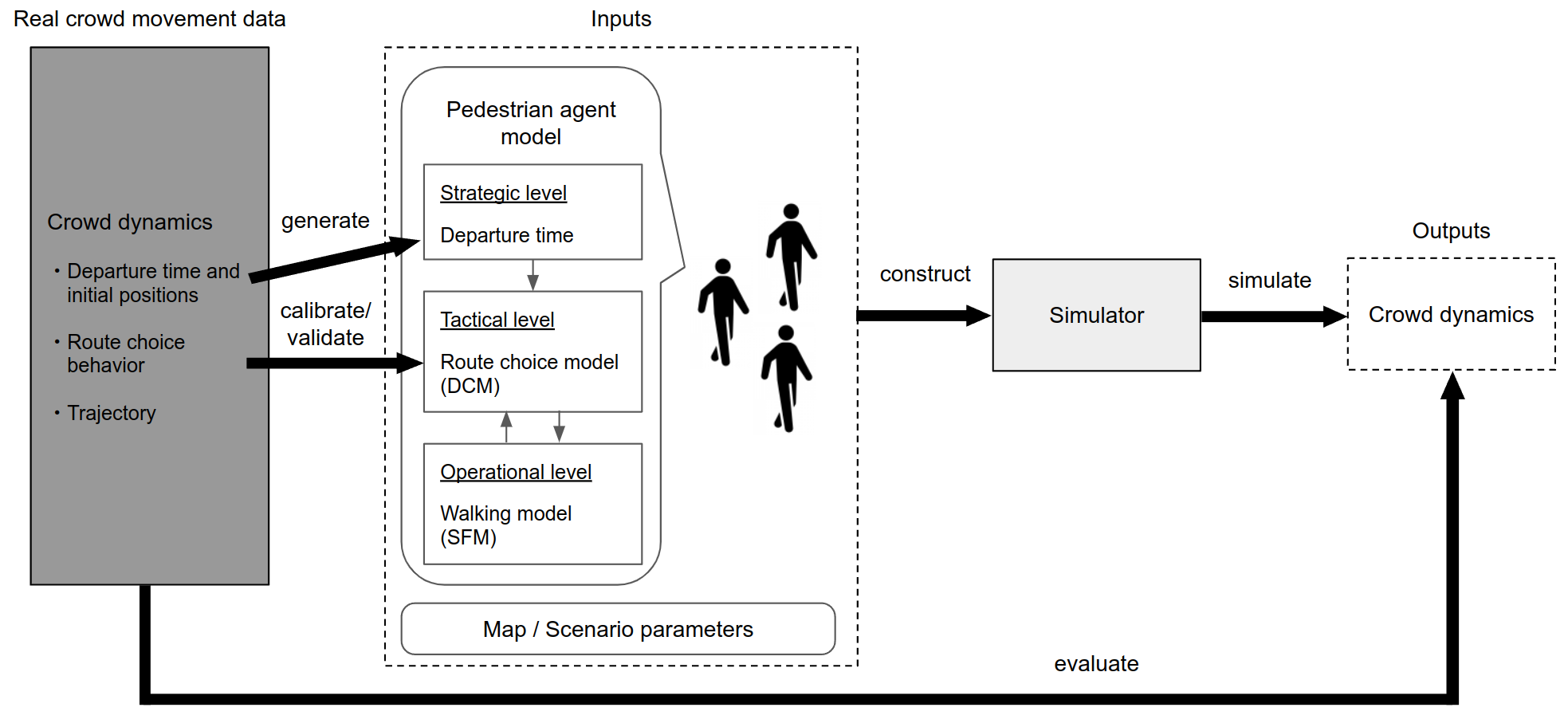}
\caption{Our crowd simulation framework}
\label{fig:overview}
\end{figure*}

Therefore, in this study, we generalize and propose a crowd simulation framework that includes actual crowd movement measurements, route choice model estimation, and crowd simulator construction. Furthermore, we verify the reproducibility of the crowd simulation incorporating the route choice model using the measured real large-scale crowd movement data.

\section{Methodology}

Our crowd simulation framework includes actual crowd movement measurements, route choice model estimation, and crowd simulator construction. The crowd simulator is agent-based and models pedestrian route choice and walking behavior. In this section, we describe the framework overview, route choice model, walking model, measurement methods, and evaluation methods.

\subsection{Simulation framework overview}

We define our crowd simulation framework as shown in Figure~\ref{fig:overview}. Measure the crowd dynamics, such as trajectory and number of pedestrians, and calibrate the model based on measured data of crowd movement. Then, a simulator is constructed based on the pedestrian agent model, the map, and other parameters, and run the simulation to output the crowd dynamics. Finally, the simulation is evaluated by comparing the actual crowd movement with the simulation results.

The pedestrian behavior can be categorized into {\it Strategic}, {\it Tactical}, and {\it Operational levels}. The model of the {\it Operational level} is mature, and the modeling of the {\it Tactical} and {\it Strategic levels} should be promoted. In this study, we focus on the modeling of route choice behavior at the {\it Tactical level}. Therefore, in our crowd simulation, the model represents the crowd movement after the decision at the {\it Strategic level} is made.
In other words, the departure time of pedestrian movement is already determined \footnote[1]{It could be that constructing a departure time choice model similar to the route choice model using DCM and generating the departure time according to the model.}. In summary, pedestrian agents are generated according to pre-defined departure time, then move according to the route choice model and the walking model.

\subsection{Route choice model}\label{rcm}
We apply DCM as the route choice model. DCM is based on random utility maximization theory \cite{McFadden1973-ut}. This theory assumes that when people make a choice, they choose the option that maximizes their utility. 
Utility essentially consists of attributes of alternatives $\bm{x}$ such as travel distance, and preferences $\bm{\beta}$ for those factors. The utility function consists of two terms. The first is a deterministic term $V$ that function of $\bm{x}$ and $\bm{\beta}$. The second is a probability term $\varepsilon$. 
The probability term is used to represent utility probabilistically since the true utility of the decision-maker is unknown to the analyst.
The utility of each route and the choice probability are calculated, and the final selection is determined. Models that assume a Gumbel distribution for the probability term are called logit models, and the most basic model is the multinomial logit model (MNL) \cite{McFadden1973-ut}. Note that in this paper, when we use the term DCM, we mean MNL.

In DCM, the utility function $U_{ij}$ for individual $i$ choice $j$ is shown below. The utility function consists of a deterministic term $V_{ij}$ and a probability term $\varepsilon_{ij}$. In general, the deterministic term is expressed as a linear sum of the $k$ observable factors $x_{ij,k}$ and preferences $\beta _{k}$. 

\begin{eqnarray}\label{utility}
U_{ij} &=& V_{ij} + \varepsilon_{ij} \\
 &=& \beta_{ij,0} + \sum_{k=1} \beta _{ij,k} x_{ij,k} + \varepsilon_{ij}
\end{eqnarray}

where $\beta_{ij,0}$ is an alternative-specific constant (ASC) that expresses a bias toward an alternative $j$. 

Assuming a Gumbel distribution for the probability term, the probability that an individual $i$ will choose option $j$ from a choice set $C$ is as follows:

\begin{eqnarray}\label{choice_prob}
P_i(j) = \frac{\exp(V_{ij})}{\sum_{l\in C} \exp(V_{il})}
\end{eqnarray}

The preference parameters $\beta_{k}$ are estimated by maximizing the log-likelihood function given by

\begin{eqnarray}\label{}
LL(\bm{\beta}) = \sum_i \sum_j y_{ij}\log P_i(j) 
\end{eqnarray}

where $y_{ij}$ is 1 when an individual $i$ chooses option $j$, and 0 otherwise. 

The model builder can list the elements $\bm{x}$ involved in the route choice and define a utility function. Then, by measuring the variable $\bm{x}$ and the route choice result $y$, the route choice model, in other words, the parameters $\bm{\beta}$ of DCM can be calibrated. The variable $\bm{x}$, which is involved in route choice, varies from scenario to scenario and is discussed in detail below.

\subsection{Walking model}
We apply SFM as the walking model. SFM is a physical model that treats pedestrians as a point mass. The forces required to go to a destination and received from other pedestrians and the environment are defined as Social Force, and the movement of pedestrians is formulated by these forces. In our simulations, we compute the Social Forces using the route direction determined by the route choice model as the destination. The model at the operational level does not have to be SFM; for example, RVO can be used. The operational level is responsible for local navigation, while the route choice model at the tactical level is responsible for global navigation. Therefore, any model that has such a function can be substituted. 

\subsection{Crowd movement measurement methods}\label{measurement_method}
We describe a method for measuring crowd movement. The departure time, initial position, and route choice behavior of each pedestrian in a crowd can be calculated by measuring the trajectory of the pedestrians. Therefore, we describe a method for measuring pedestrians' trajectories.

We introduce the measurement methods using an RGB-depth camera or LiDAR.
These sensors can measure point cloud information of three-dimensional objects such as pedestrians. The point clouds are clustered, and clusters with a large number of points are identified as pedestrians. The highest point in the cluster is the pedestrian's head, and the trajectory is extracted by discriminating that point in the time series \cite{Onishi2015}. 

The RGB-depth camera can extract the trajectories with a small error margin in bright indoor environments, but it has the disadvantage that it cannot be used outdoors with the sun since it actively emits infrared rays. On the other hand, LiDAR can detect pedestrians even outdoors since it measures the distance to surrounding objects from the time it takes for a laser beam to hit an object and bounce back, and noise related to sunlight can be cut. 
The disadvantage of LiDAR is that the laser is blocked in bad weather conditions, such as rain or snow, degrading the measurement performance.


\subsection{Evaluation}
This section describes the evaluation of the route choice model and the evaluation of the crowd simulation.
We can use the evaluation method for general classification problems, such as in machine learning, for the evaluation of the route choice model itself. We separate the choice behavior data for training and test data and evaluate the choice prediction accuracy of the test data. 

The number of people who choose the route is compared between the actual and the simulation for the evaluation of the crowd simulation. This is a reasonable evaluation index when focusing only on route choice behavior. Furthermore, the important indicators in crowd movement are efficiency and safety. Efficiency is often evaluated by travel time, and safety is often evaluated by congestion degree.  
Therefore, it is also necessary to evaluate the reproducibility of the crowd simulation using indicators related to travel time, such as the number of people completing the trip at each time point.

\section{Evacuation drill}
We collected crowd route choice behavior at an evacuation drill to verify the reproducibility of the crowd simulation incorporating the route choice model, which was conducted at the New National Theater in Japan. The opera was actually performed in the evacuation drill, an evacuation order was issued during the performance, and the audience evacuated from the theater.
Figure~\ref{fig:shinkoku}(a) and (b) show the layout of the theater and the evacuation route. The red route leading straight out of the auditorium is the correct evacuation route. It is also possible to choose the blue route and proceed toward the stairs.

\subsection{Measurement}
\begin{figure}[b]
\centering
\includegraphics[width=1.0\linewidth]{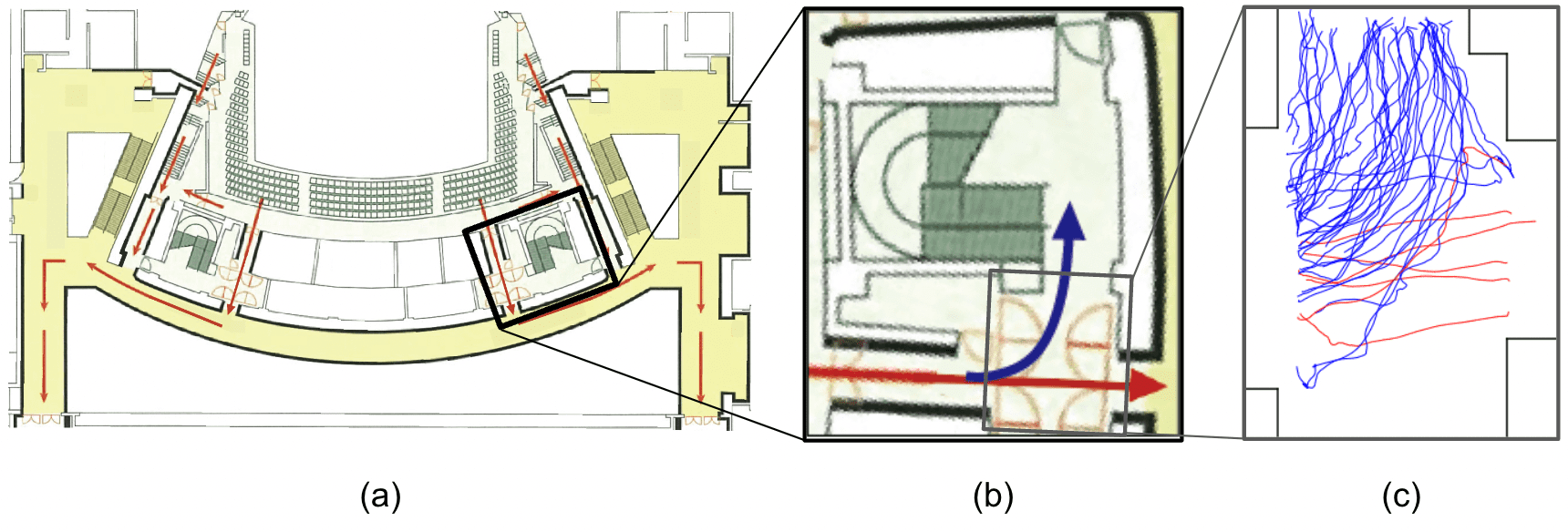}
\caption{Measurement area and data. (a) The layout of the theater (b) details of the measurement area (c) The pedestrians' trajectories. They are color-coded according to the final route chosen.}
\label{fig:shinkoku}
\end{figure}

We measured crowd movement using RGB-Depth cameras at the locations shown in Figure~\ref{fig:shinkoku}(b) where pedestrians exit the hallway from the auditorium. 52 people were evacuated through this door.
Figure~\ref{fig:shinkoku}(c) shows the trajectories extracted from the RGB-Depth camera data using the method described in Section~\ref{measurement_method}. We calculated the direction of the body per 0.5 [s] from these trajectories. Then, we obtained the pedestrian's chosen route per 0.5 [s] by defining the route in the direction of the pedestrian's body as the selected route. 

\begin{figure}[tb]
    \centering
    \subfigure[]{\includegraphics[width=0.32\linewidth]{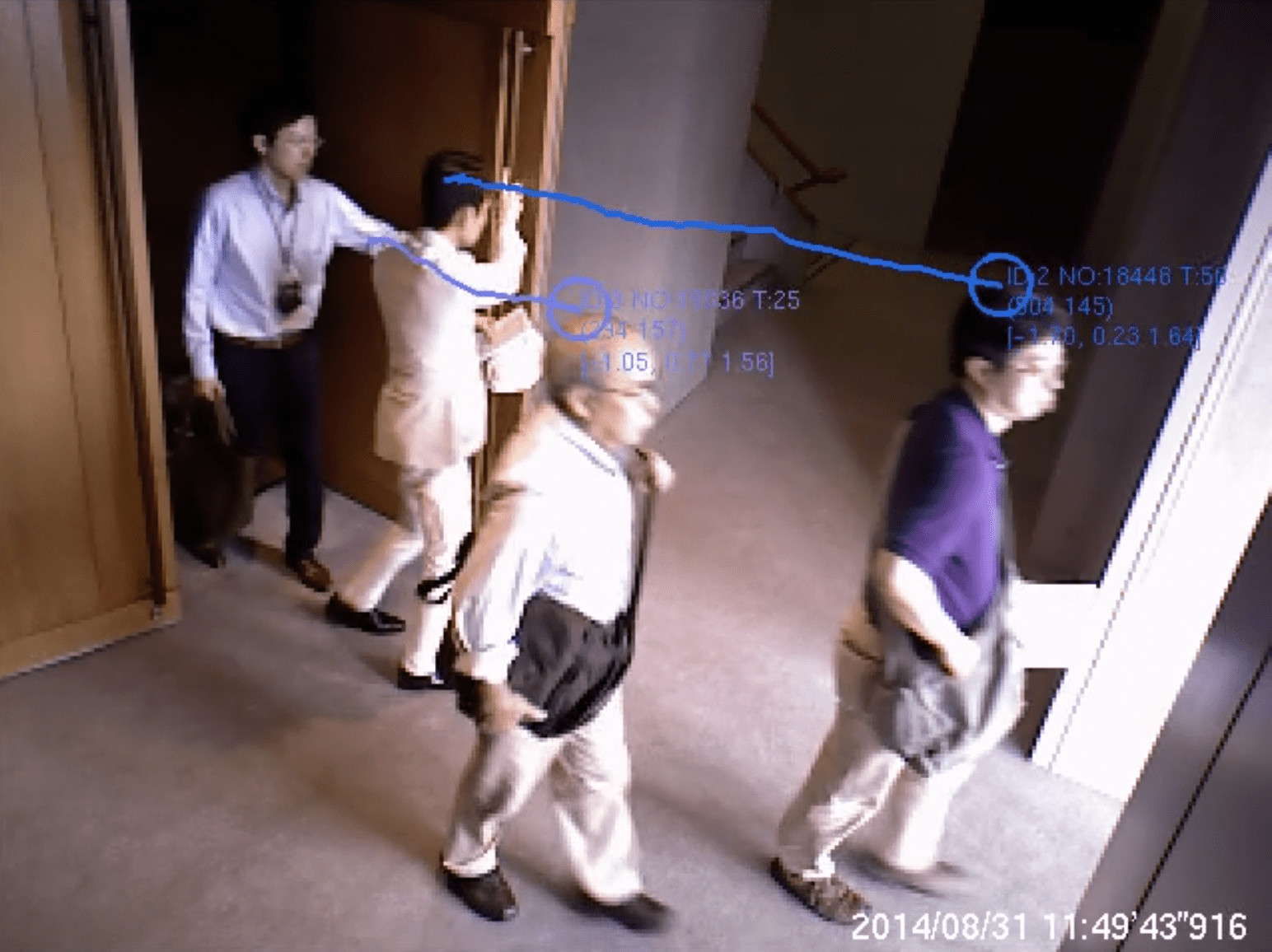}}%
    \hspace{0.05pt}
    \subfigure[]{\includegraphics[width=0.32\linewidth]{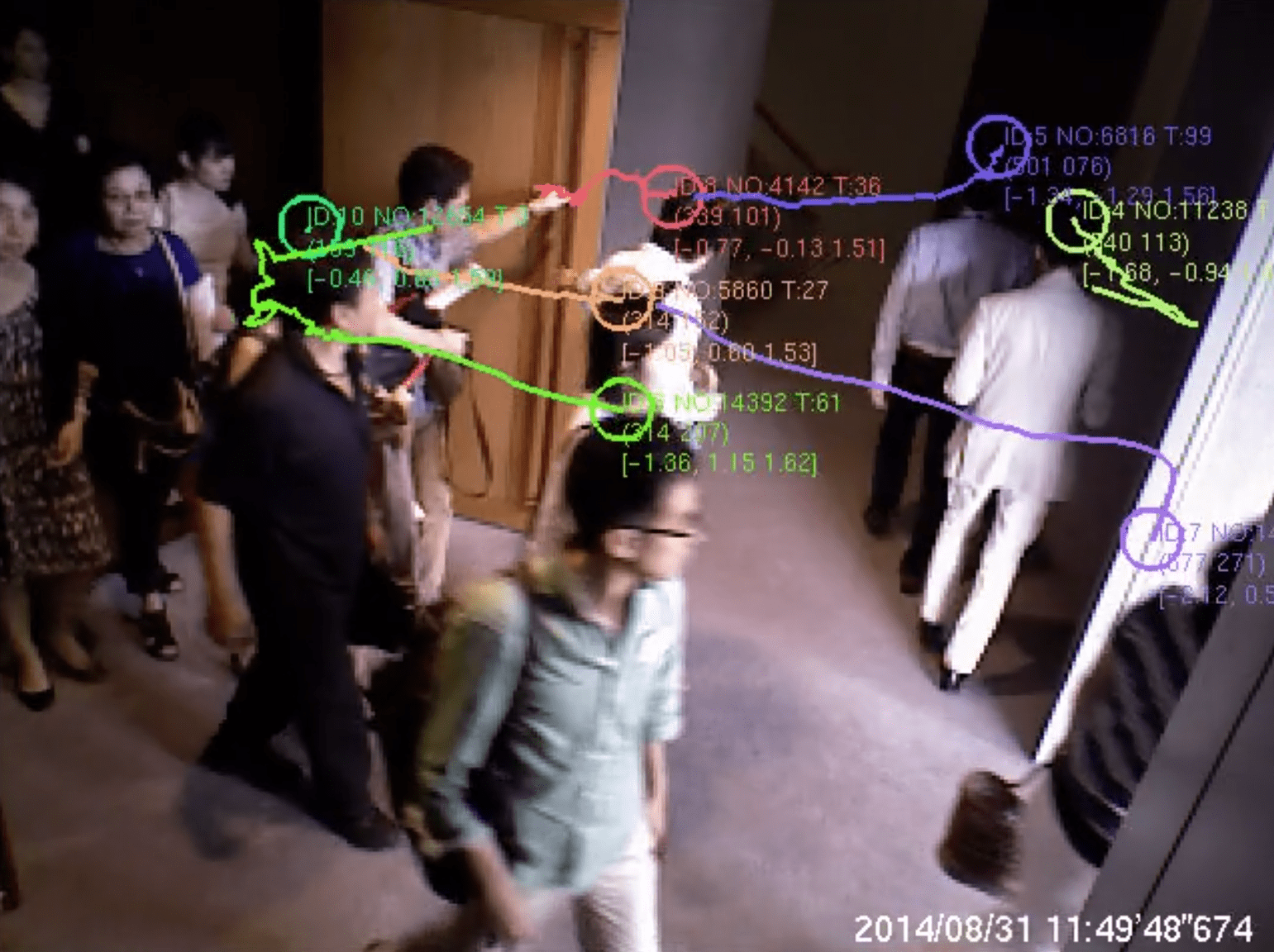}}%
    \hspace{0.05pt}
    \subfigure[]{\includegraphics[width=0.32\linewidth]{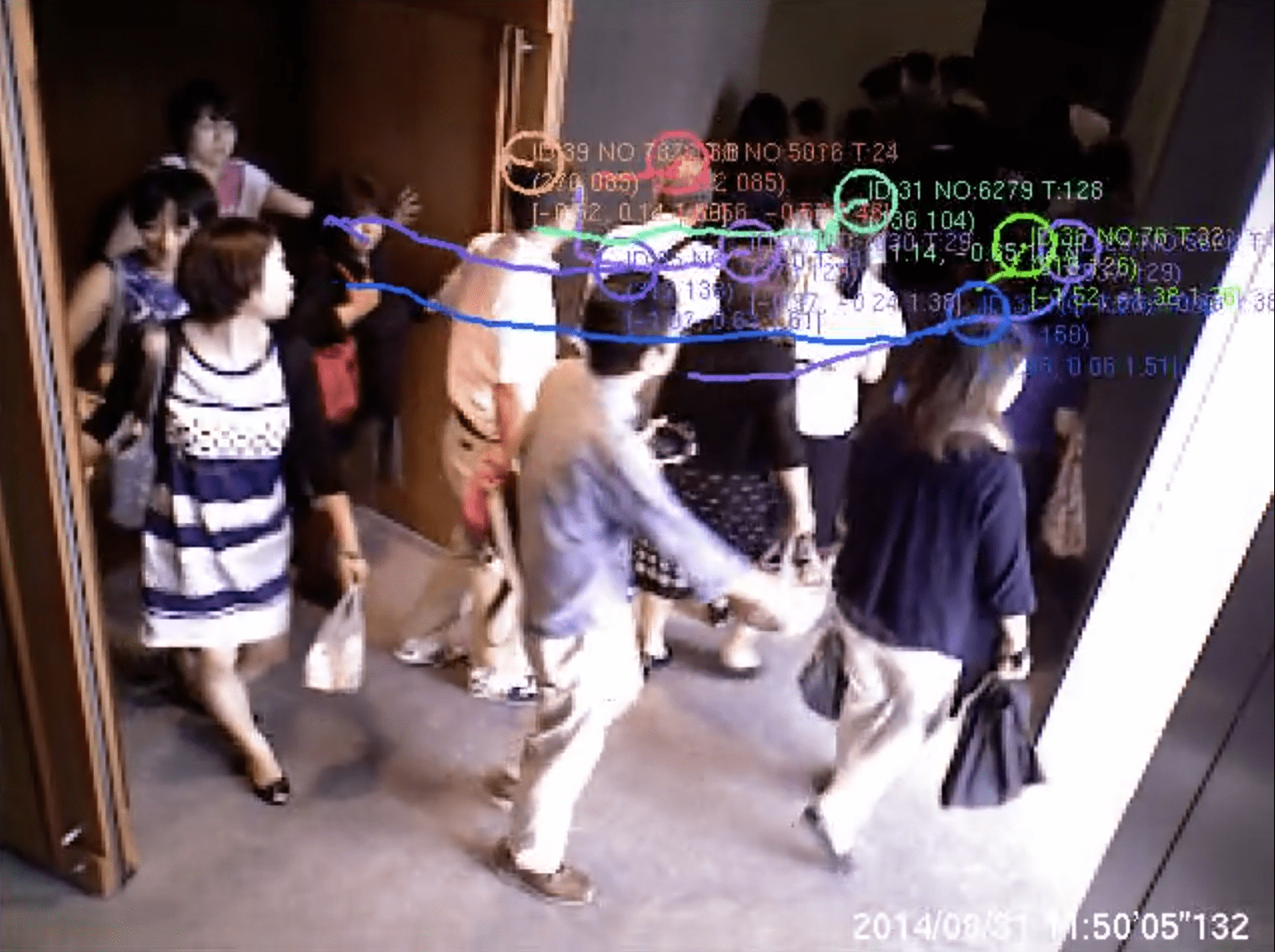}}%
    \caption{The changes in pedestrians' route choice. The flow of pedestrians changes in (a), (b), and (c).}
    \label{fig:branch2}
\end{figure}

At the locations, a phenomenon was measured in which the crowd flow changed, depending on the choice of a few evacuees, as shown in Figure~\ref{fig:branch2}. First, they went straight out the door (a), but when two people chose the stairway (b), the pedestrians in the rear also chose the route to the stairway, following their choice (c). We attempt to model such route choice behavior and simulate crowd movement.

\subsection{Modeling}
We describe a method for modeling route choice behavior using DCM. As explained in Section~\ref{rcm}, it is necessary to define the alternative attributes, i.e., the factors involved in route choice. The factors we considered are:

\begin{itemize}
    \item \textbf{DIST}ance from the position of decision-maker to the start point of the route (DIST);
    \item \textbf{CH}osen at the previous step (0.5 [s] before) or not (CH);
    \item \textbf{N}umber of pedestrians choosing the route who are in \textbf{F}ront of the decision-maker (NF);
    \item \textbf{N}umber of pedestrians choosing the route who are \textbf{B}ehind the decision-maker (NB);
\end{itemize}

\begin{figure}[tb]
    \centering
    \subfigure[DIST]{\includegraphics[width=0.28\linewidth]{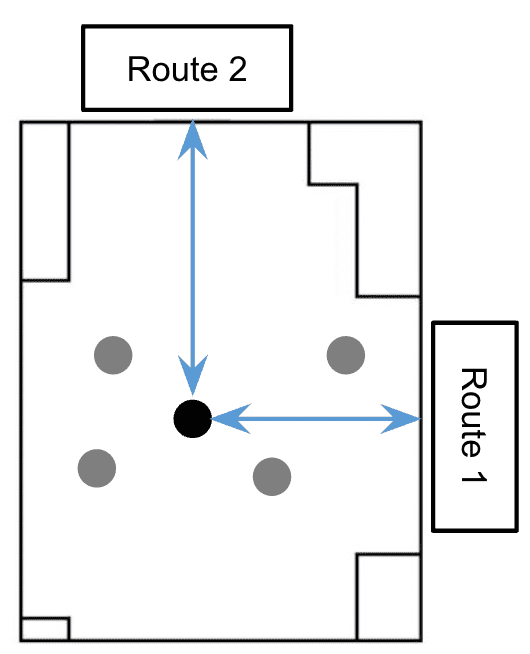}}%
    \hspace{0.05pt}
    \subfigure[CH]{\includegraphics[width=0.22\linewidth]{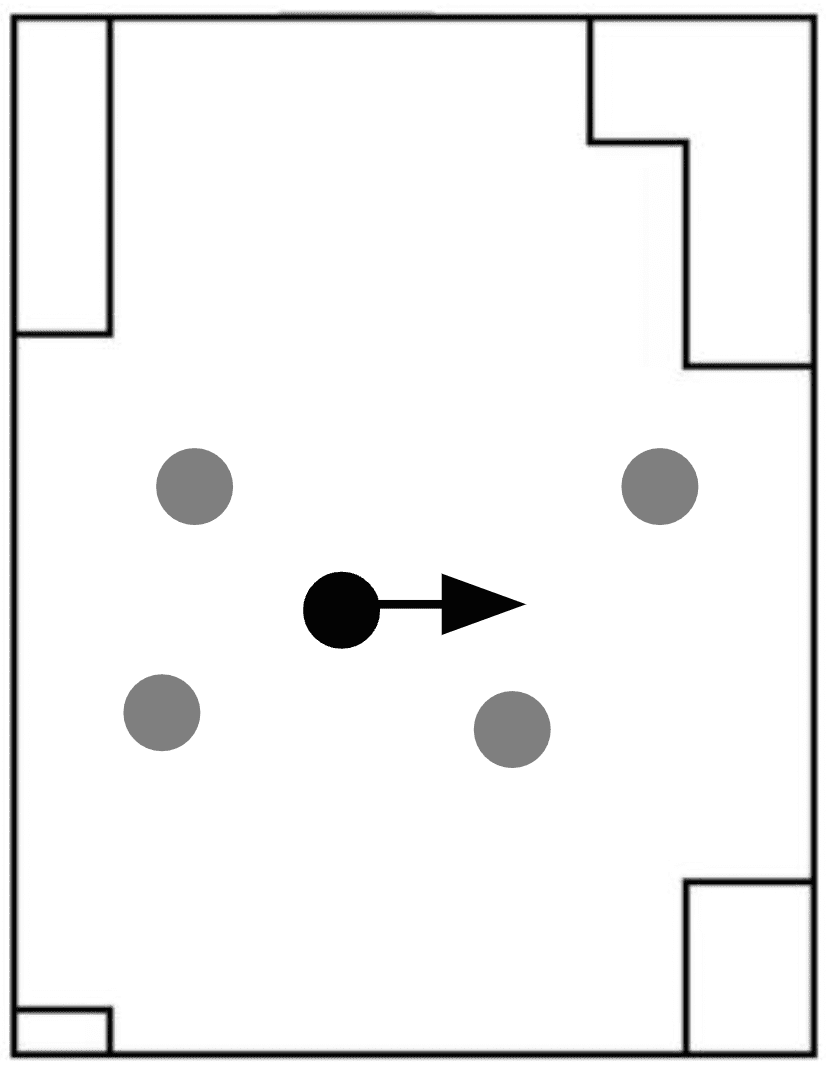}}%
    \hspace{0.05pt}
    \subfigure[NF]{\includegraphics[width=0.22\linewidth]{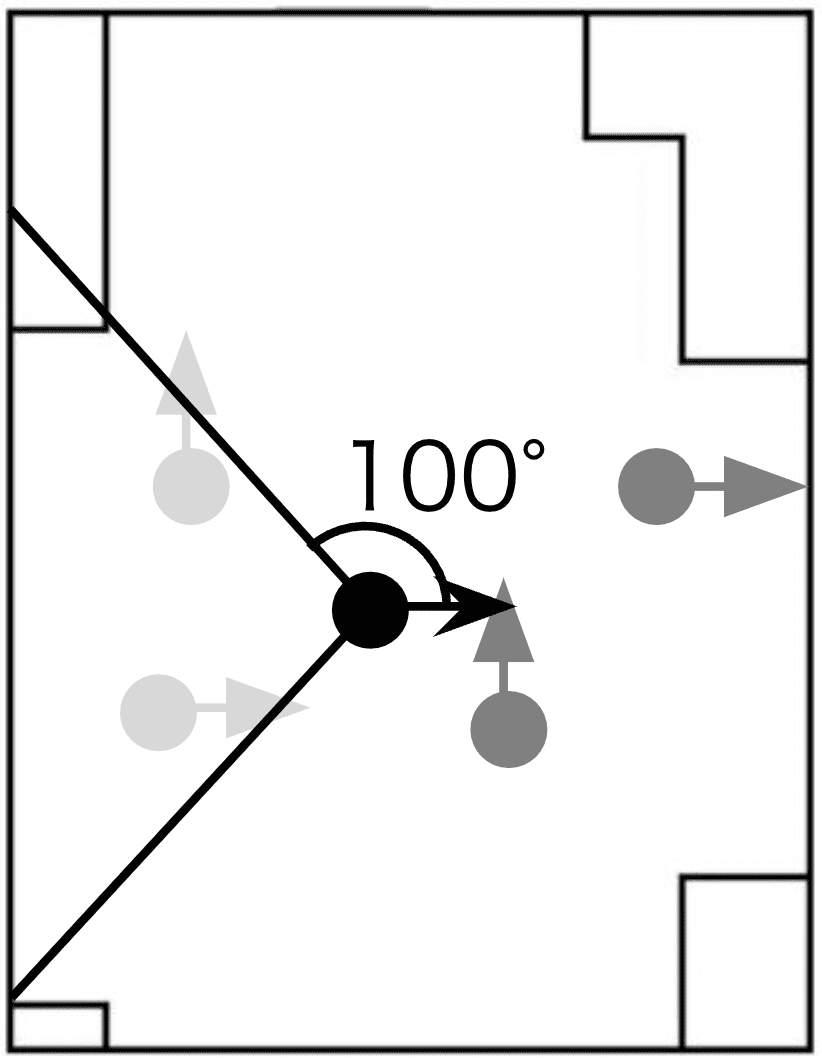}}%
    \hspace{0.05pt}
    \subfigure[NB]{\includegraphics[width=0.22\linewidth]{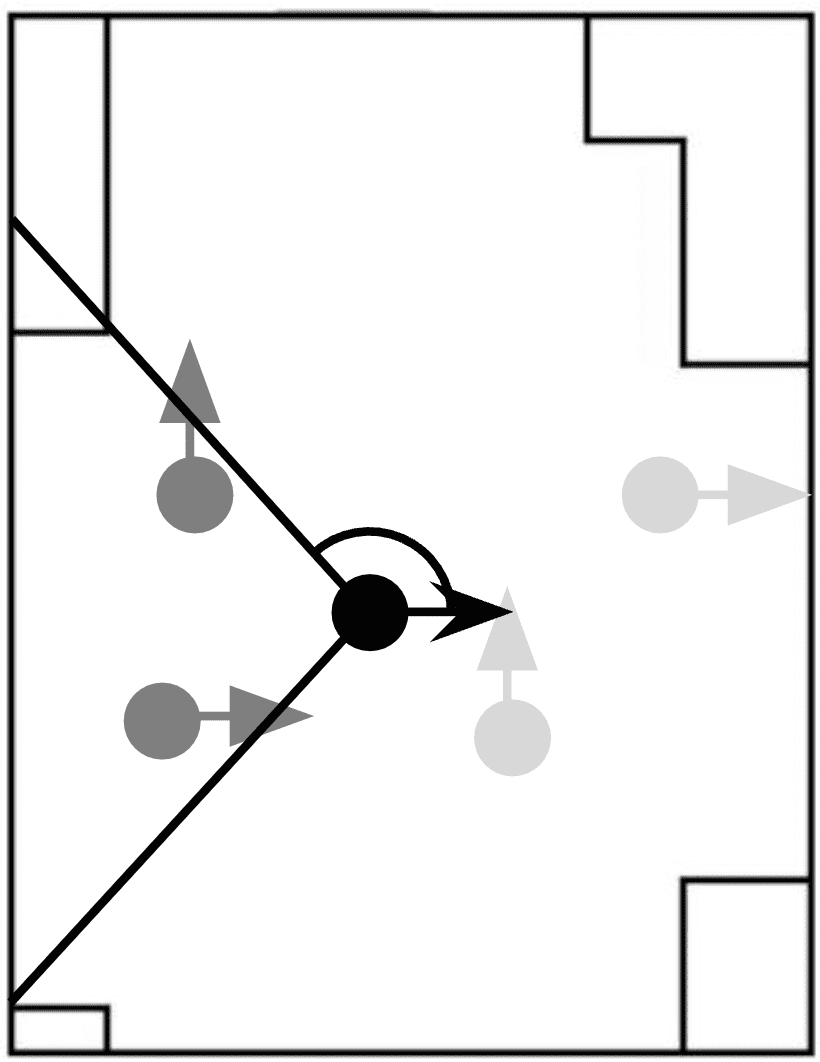}}%
    \caption{The factors involved in route choice in the evacuation}
    \label{fig:shinkoku_factors}
\end{figure}

Figure~\ref{fig:shinkoku_factors} shows the graphical explanation of the factors. In this situation, the value of each factor in Route 1 is as follows. DIST is the distance from the pedestrian's position to the starting point of Route 1. CH is the selection made in the previous step, and the value is 1 in this case because Route 1 was selected, otherwise the value is 0. NF and NB are each 1 since there is one person in the front and one person in the rear who has chosen Route 1.

The determinant term of the utility function is defined as follows. First, in this study, we change $\beta_{ij,k}$ in Eq.~\ref{utility} to $\beta_{k}$ by assuming that preference $\bm{\beta}$ does not vary from individual to individual and from option to option, so the determinant term of the utility function is as follows:

\begin{eqnarray}
V_{ij} = ASC_{j} + \sum_{k=1} \beta_{k} x_{ij,k} 
\end{eqnarray}

Therefore, the determinant term of the utility function in the case of evacuation drills is as follows:

\begin{eqnarray}
V_{ij} &=& ASC_{j} + \beta_{DIST} DIST_{ij} + \beta_{CH} CH_{ij} \\ 
&+& \beta_{NF} NF_{ij} + \beta_{NB} NB_{ij} \nonumber
\end{eqnarray}
where $j$ is Route 1 or Route 2 and $ASC_{Route 1} = 0$. Assume that for every step (every 0.5 [s]), pedestrian $i$ computes the utility of Route 1 and Route 2 and chooses a route.

\begin{table}[tb]
    \centering
    \caption{Estimation result of DCM in the evacuation drills}
    \input{tab/shinkoku_model_parameters_v}
    \label{tab:shinkoku_model}
\end{table}

We divide the measurement data into 5 parts per pedestrian unit, and estimate the route choice model through a 5-fold cross-validation. Table~\ref{tab:shinkoku_model} lists the estimated parameters of the model and the prediction accuracy. The estimated parameters can be used to understand the route choice behavior of pedestrians. First, the negative value of $\beta_{\rm DIST}$ indicates that pedestrians are less likely to choose a route with a long distance to the starting point of the route. The positive value of $\beta_{\rm CH}$ indicates that pedestrians are more likely to make the same choice as the previous step. 
The fact that $\beta_{\rm NF}$ is positive indicates that the decision-makers tend to choose the route with the greater number of choosers among the number of pedestrians in front of them. In other words, it indicates that the decision maker follows other pedestrians and it strengthens the evidence for the existence of herding behavior. 
Haghani et al. reported the existence of herding behavior in an empirical experiment imitating an evacuation situation \cite{Haghani2017-bv}, and our study supports that claim with data from more realistic evacuation situations.
Interestingly, $\beta_{\rm NB}$ is negative and half the value of $\beta_{\rm NF}$. This indicates that the choice of the pedestrian behind the decision maker has less impact on that decision maker than the choice of the pedestrian in front of the decision maker. Rather, it indicates that the decision-makers are more likely to choose the opposite direction to the route chosen by most of the pedestrians behind them.

\subsection{Simulation}
We test whether a crowd simulation consisting of the estimated route choice model (DCM) and the walking model (SFM) can represent crowd movement during an evacuation drill. We use actual data of the pedestrians' initial positions and the time they exit from the door. 
The first pedestrian to exit the door and the two pedestrians who chose Route 2 that triggered the change of the crowd flow are assumed to move in the simulation as they did in the actual situation. All other pedestrians move according to DCM and SFM after they are generated according to the actual initial position and timing. The SFM parameters are the same as in \cite{Helbing1995}.

In this experiment, we evaluate the reproducibility of the simulation by the number of people selecting the route. The shortest path selection model (SP) is used for comparison. The pedestrian's route choice is stochastic, so the simulation is run 50 times.

\begin{figure}[tb]
\centering
\includegraphics[width=1.0\linewidth]{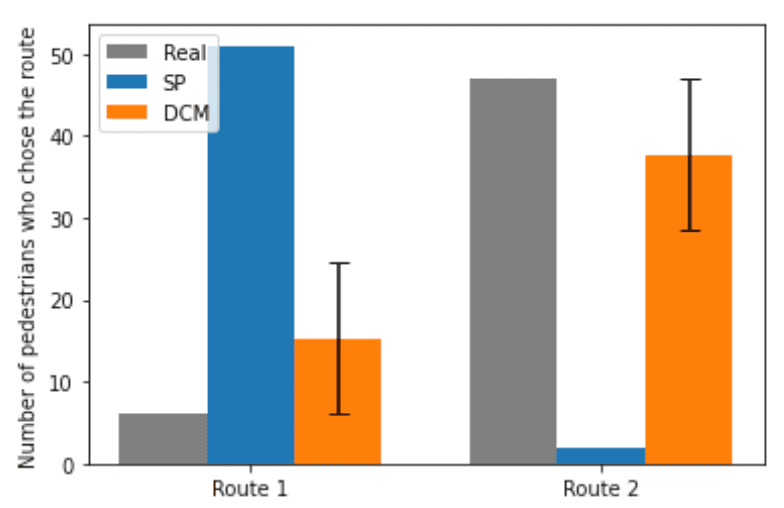}
\caption{Number of pedestrians who chose the route}
\label{fig:result_shinkoku}
\end{figure}

Figure~\ref{fig:result_shinkoku} illustrates the number of people selected for Route 1 and 2. 
Most pedestrians choose Route 1 when the route choice model is SP. On the other hand, DCM takes into account the choices of other pedestrians’ choices in addition to distance. Therefore, the influence of the choice of the pedestrian who triggered the change in the flow of the crowd to Route 2 can be considered. Consequently, more pedestrians choose Route 2. 
Route selection is stochastic, and the final route they choose will change even if pedestrians start moving under the same conditions. Therefore, the results of the simulation vary, but the actual crowd movement is one of the results of the crowd movement simulated by DCM.

\section{Firework event}
In this section, we verify the reproducibility of the crowd simulation incorporating the route choice model using more large-scale crowd movement.
During mass gathering events such as the haji, football matches, music festivals, and firework events, crowd control is necessary \cite{Soomaroo2012}. Crowd simulation is an imperative tool to test and improve crowd control strategies, and it is required to verify the crowd simulation using crowd movement in such mass gathering events.  

The target of this study is crowd movement at the firework event held at Moji Port in Kitakyushu, Japan.
Tens of thousands of people move from the event site to the nearest station as the end of the fireworks display approaches. Security guards are in charge of guiding and controlling the crowd at several points to avoid the risk of congestion and accidents, which occur if many people flow into the station simultaneously.

\begin{figure}[tb]
\centering
\includegraphics[width=1.\linewidth]{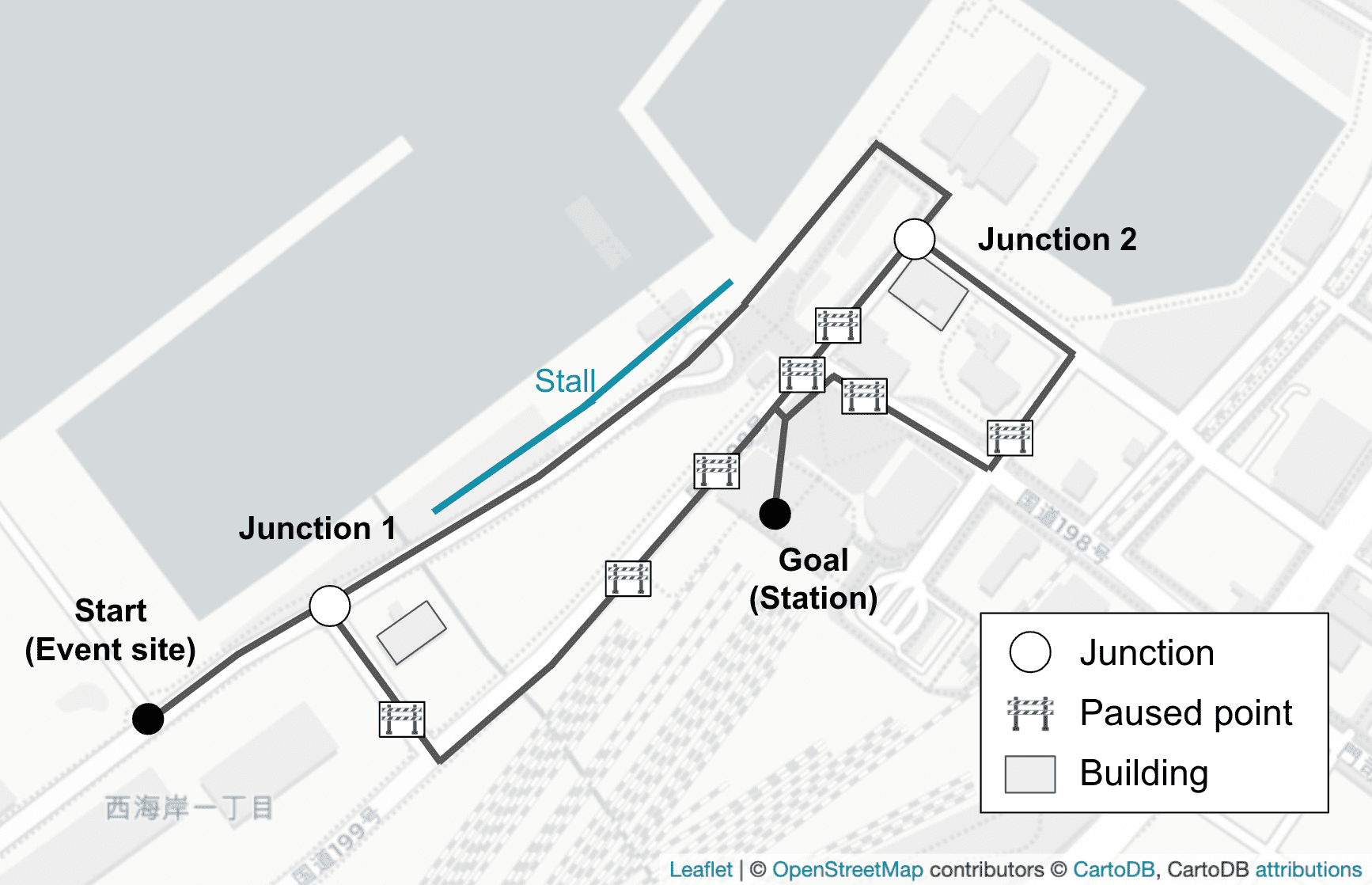}
\caption{Routes and guidance control points in the firework event}
\label{fig:moji}
\end{figure}

Figure~\ref{fig:moji} shows the route from the event site to the station and the control points. There are two junctions and seven paused points. At the junctions, route guidance control is provided for going straight or detouring, and guidance control is provided for proceeding or stopping for a certain period of time at the paused points.
At the junction, in addition to the guidance control by security guards, guidance information is presented by guide projection on the building. In addition, food stalls are open on the route straight ahead from Junction 1.

\subsection{Measurement}
We measured route choice behavior at junctions and the number of pedestrians arriving at the station. In addition, we measured the control at each control point. Here, we describe each of the measured information.

\begin{figure}[tb]
    \centering
    \subfigure[Measurement of the pedestrian movement with LiDAR ]{\includegraphics[width=0.5\linewidth]{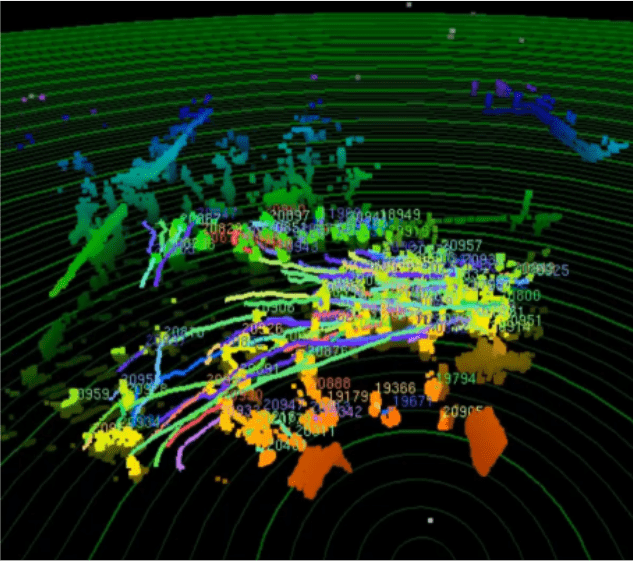}}%
    \hspace{0.05pt}
    \subfigure[Trajectories ]{\includegraphics[width=0.46\linewidth]{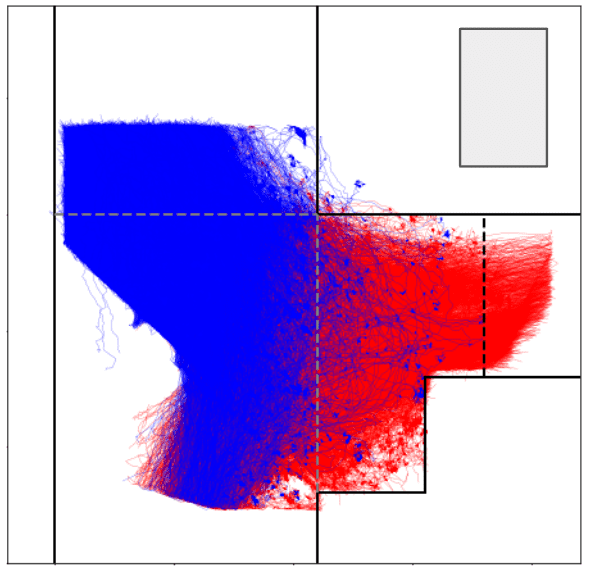}}%
    \caption{Measurement at Junction 1}
    \label{fig:junction1}
\end{figure}

First, we used LiDAR to measure route choice behavior at junctions. Figure~\ref{fig:junction1}(a) shows the state of measurement with LiDAR at Junction 1. We extracted the trajectories from LiDAR's data which is point cloud information, as shown in Figure~\ref{fig:junction1}(b). Then, identify the chosen route at the junction from the trajectory data. Here, instead of identifying a chosen route every 0.5 [s], as in Section 4, we extracted a one-time route choice at the two junctions. As a result, a total of 34937 route choice behaviors at Junctions 1 and 2 were collected.

\begin{figure}[tb]
    \centering
    \subfigure[Measurement of the pedestrian movement with RGB-Depth camera ]{\includegraphics[width=0.39\linewidth]{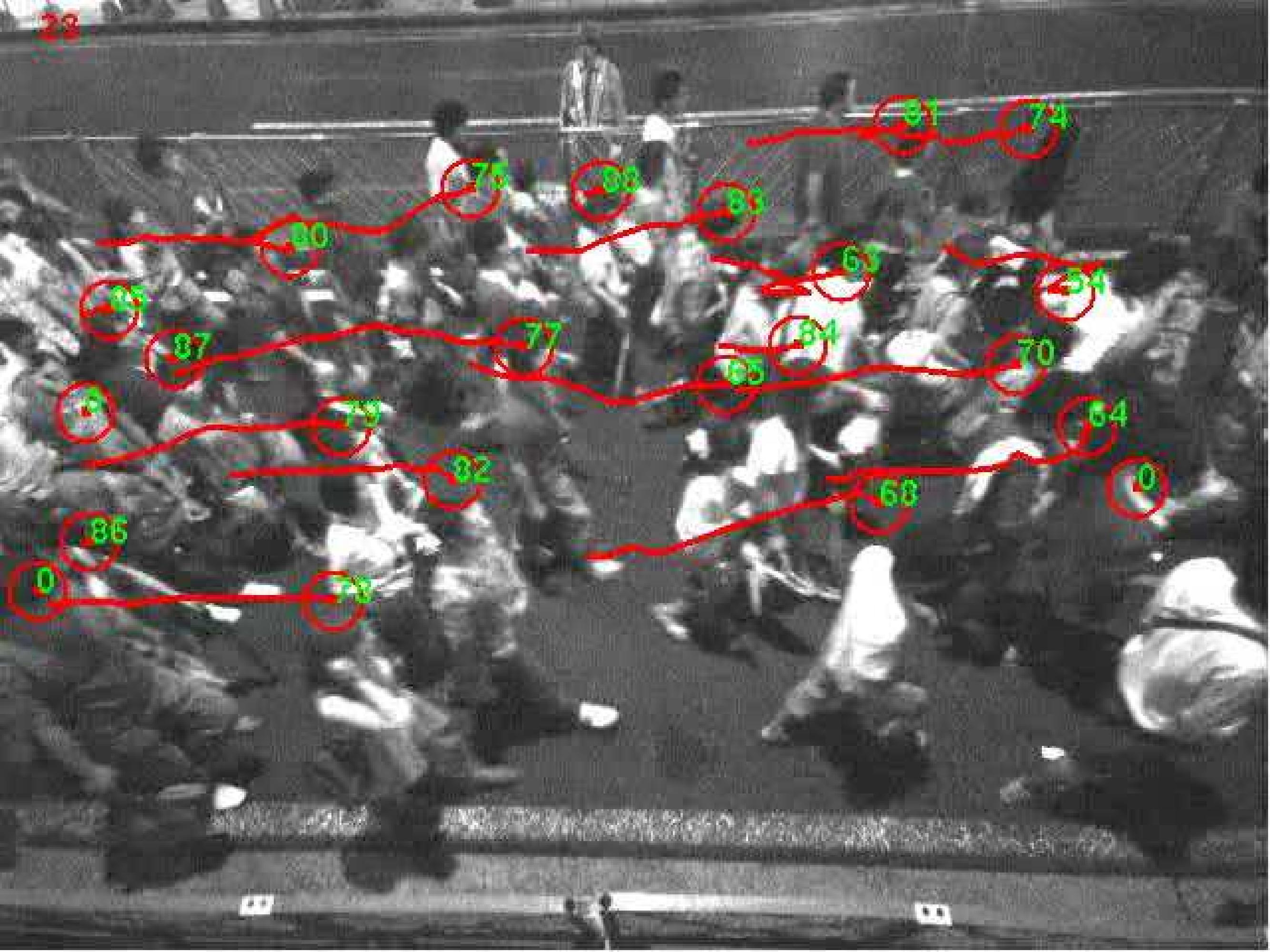}}%
    \hspace{0.3pt}
    \subfigure[The number of people arriving at the station at each time]{\includegraphics[width=0.59\linewidth]{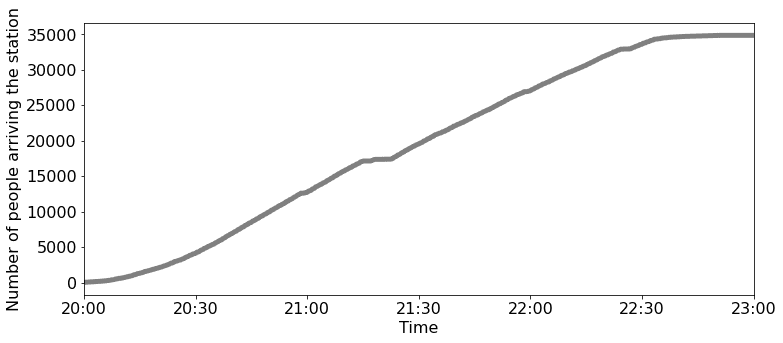}}%
    \caption{Measurement at the station}
    \label{measure}
\end{figure}

Second, the RGB-Depth cameras were also installed at the station to count the number of pedestrians arriving at the station, as shown in Figure~\ref{measure}(a), and Figure~\ref{measure}(b) shows the result. 
The total number of people is 34839. The fireworks show ended at 20:40, but people started returning to the station at 20:00 and arrived at the station by 23:00.

\begin{figure}[tb]
\centering
\includegraphics[width=1.\linewidth]{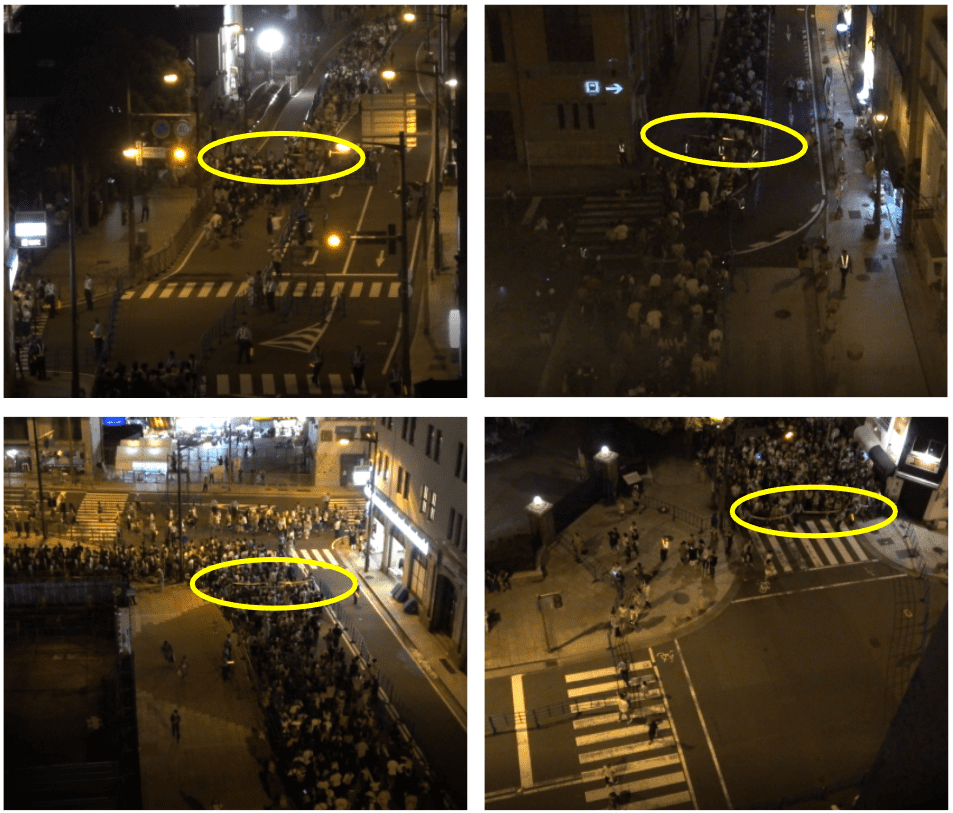}
\caption{Guidance at each control point}
\label{fig:guide}
\end{figure}

Finally, we installed cameras at junctions and paused points and measured the guidance that was actually carried out, as shown in Figure~\ref{fig:guide}. The guidance was switched according to the situation by the guards on the point.

\subsection{Modeling}

We describe a method for modeling route choice behavior. Note that we assume that the pedestrian makes a route choice only once at a junction. The factors we considered are:

\begin{itemize}
    \item \textbf{DIST}ance of the decision-maker from the junction to the station (DIST);
    \item \textbf{GUIDE}ance of the route (GUIDE);
    \item \textbf{ATT}raction of the route, such as stall (ATT);
\end{itemize}

\begin{figure}[tb]
\centering
\includegraphics[width=1.\linewidth]{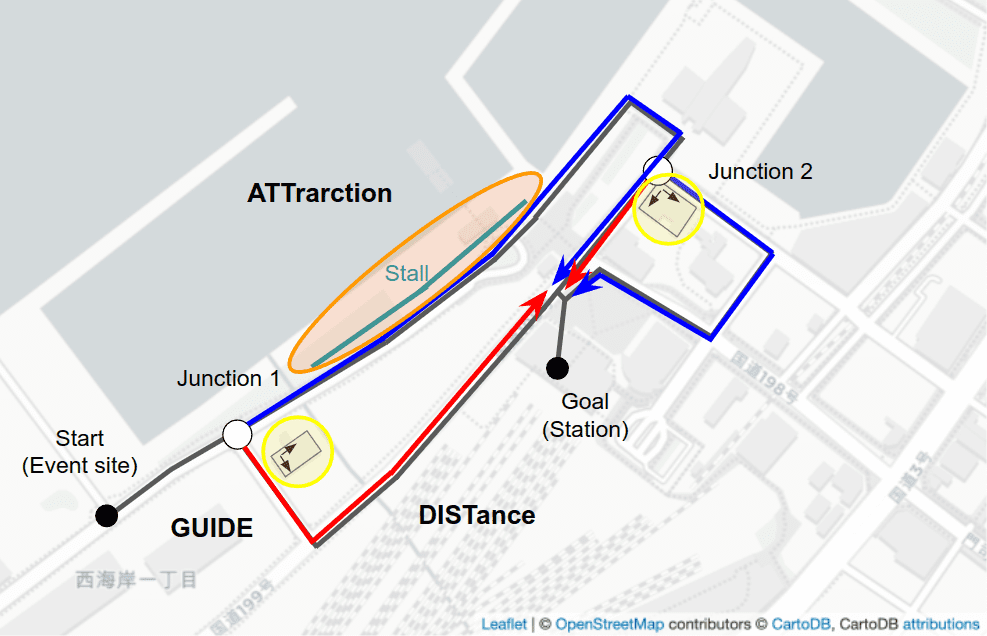}
\caption{The factors involved in route choice in the firework event}
\label{fig:moji_factors}
\end{figure}

Here, we define that the route closer to the station (red route in Figure~\ref{fig:moji_factors}) is Route 1, and the other route (blue route) is Route 2 at each junction. The distance to the station on each route at each junction is one of the factors. In addition, the presence or absence of route guidance at a junction is also considered a factor related to route choice. If there is an induction, set it to 1, otherwise 0. And we assume that the attraction of the route is influenced by the presence or absence of food stalls. If there is a stall on the route, it is set to 1, otherwise 0. Note that the stall close after 22:00.

Based on the above, the determinant term of the utility function in the case of the firework event is as follows:

\begin{eqnarray}
V_{ij} &=& ASC_{j} + \beta_{DIST} DIST_{ij} + \beta_{GUIDE} GUIDE_{ij} \\
&+& \beta_{ATT} ATT_{ij} \nonumber
\end{eqnarray}
where $j$ is Route 1 or Route 2 and $ASC_{Route 1} = 0$.

We divide the measurement data into 5 parts, and estimate the route choice model through a 5-fold cross-validation, as in Section 4. Table~\ref{tab:moji_model} lists the estimated parameters of DCM and the prediction accuracy. The following can be deduced from the estimated parameters. First, pedestrians are less likely to choose a route with a long distance. Second, pedestrians tend to choose guided routes and routes with stalls. 

\begin{table}[tb]
    \centering
    \caption{Estimation result of DCM in the firework event}
    \input{tab/moji_model_parameters}
    \label{tab:moji_model}
\end{table}

\subsection{Simulation}

\begin{figure}[tb]
\centering
\includegraphics[width=1.\linewidth]{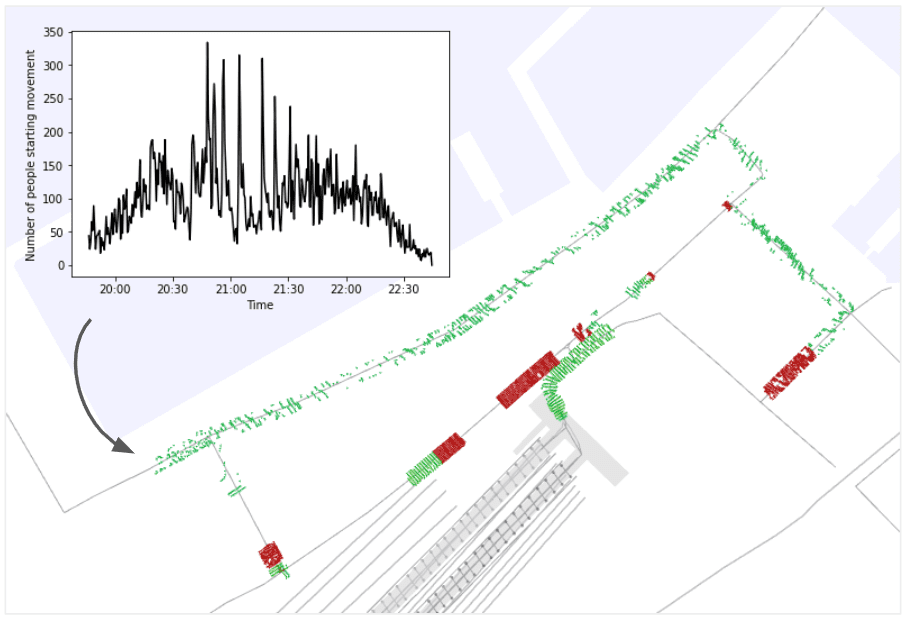}
\caption{Simulation input and visualization. The dots represent pedestrians, and their color represents the pedestrian's speed. The speed decreases from green to red, and red means that the pedestrian is stationary at zero speed.}
\label{fig:sim}
\end{figure}

We test whether a large-scale crowd simulation consisting of the estimated DCM and SFM can represent crowd movement during the firework event. We create the pedestrians' departure time from the event site and the number of people as follows. 
First, the number of pedestrians measured at Junction 1 at each time point measured by LiDAR is used as the base. 
Then, we multiply a constant by it so that its sum matches the total number of people measured at this firework event, 34839. The graph in Figure~\ref{fig:sim} shows the number of departure pedestrians at each time. In the simulation, agents are generated at the starting point according to this distribution. 
Then, pedestrian agents select a route at junctions 1 and 2 according to DCM, and walk on the route according to SFM. 
Guidance at junctions and stops is performed as the actual measured operation in the simulation.
At the station, trains operate according to the actual timetable. The capacity of the trains is fixed, and pedestrians who cannot board a train wait at the station until the next train arrives.

We use CrowdWalk, an open-source multi-agent pedestrian simulator \footnote[2]{https://github.com/crest-cassia/CrowdWalk}, as a base simulator \cite{Yamashita2013-yj}. CrowdWalk expresses the movable area using one-dimensional links and nodes, and it is light in memory consumption and calculation time. Therefore, CrowdWalk is suitable for large-scale simulations including tens of thousands of agents. CrowdWalk uses SFM as a walking model of pedestrian agents. The default parameters of SFM of CrowdWalk are used in this experiment. The default route choice of agents in CrowdWalk is SP. Therefore, we extend the functionality of CrowdWalk to allow for utility calculation and DCM-based route choice. Figure~\ref{fig:sim} shows the visualization of crowd simulation using CrowdWalk.

In this experiment, we evaluate the reproducibility of the simulation by the number of people arriving at the station at each time point. 
The evaluation index is MAE (Mean Absolute Error) and RMSE (Root Mean Square Error) of the real data and simulation results. RMSE tends to treat outliers (large deviations) as larger errors than MAE. Therefore, we use both as evaluation indexes. 
For comparison, the reproducibility of a simulation in which the pedestrian perfectly follows the guidance is also calculated.
We call the simulation setting as Follow. In addition, we compare the effect on computation time of adding route choice to the agent's model in a crowd simulation of tens of thousands of people. We use Intel(R) Core(TM) i9-9900K CPU (3.60GHz) to run the crowd simulation. In the case of using DCM as the route choice model, the pedestrian's route choice is stochastic, so the simulation is run 50 times.

Table~\ref{tab:moji_results_tab} lists the performance of the crowd simulation. 
The value for DCM represents the average and standard deviations of each metric. 
It can be seen that the use of DCM improves the reproducibility compared to Follow case, which did not consider pedestrians' decision-making. Compared to Follow, the use of DCM improved the error by 22.3\% for MAE and 15.6\% for RMSE. In addition, the computation time of the simulation with DCM is 1.13 times longer than Follow, which does not model route choice, but this is acceptable for improving reproducibility. 

\begin{figure}[tb]
\centering
\includegraphics[width=1.\linewidth]{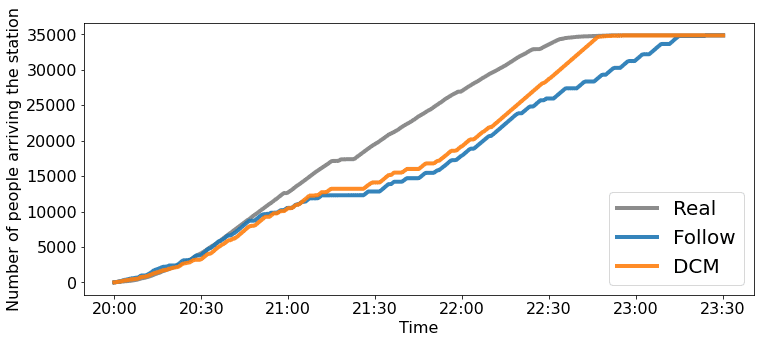}
\caption{The number of people arriving at the station at each time.}
\label{fig:moji_results}
\end{figure}

Figure~\ref{fig:moji_results} shows the distribution of the number of people arriving at the station at each time, where DCM is the average of 50 times. 
In the case of Follow, all pedestrians do not arrive at the station until after 23:00, while in the case of DCM, all pedestrians have completed their movements at 23:00, as in the actual case. 
The reason why the number of people arriving at the station in the DCM and Follow cases remained the same until about 22:30 is that the number of people waiting for the train to arrive at the station accumulated, and the congestion spread outside the station, making it difficult for pedestrians to arrive at the station. This difference from reality may stem from the expressiveness of the pedestrian behavior at the stations in the simulator.
However, in the case of DCM, pedestrians sometimes wait until the guidance changes to take the shorter route to the station since pedestrians choose their route not only based on guidance but also on the distance to the station and the attraction, such as the food stalls. As a result, more pedestrians are heading to the station more quickly in the DCM case, allowing them to arrive at the station more smoothly after the congestion at the station is reduced. Therefore, the DCM case completes all pedestrian movements faster than Follow, as in reality.

\begin{table}[tb]
    \centering
    \caption{The performance of the crowd simulation}
    \input{tab/moji}
    \label{tab:moji_results_tab}
\end{table}

\section{Conclusion}
In this study, we proposed and evaluated the crowd simulation incorporating the route choice model based on real crowd movement data. In experiments, we measured the crowd movements during the evacuation drill in the theater and the firework event in which tens of thousands of people moved. The results showed that the reproducibility of crowd simulation can be improved by introducing a utility-based route choice model. 

To our knowledge, this is the first time that the modeling and simulation construction, and the reproducibility evaluation of the simulation using large-scale real crowd movement data.
However, we did not investigate whether the constructed crowd simulations can be used to predict unknown crowd movements that were not used in the modeling. For example, we did not investigate how well it could reproduce evacuation movements in other areas of the theater or crowd movements at the next year's firework event. We plan to continue measurements and conduct such an investigation in the future.

A limitation of our approach is that it requires detailed route choice behavioral data for estimating the route choice model. However, it is difficult to obtain such data, and it is simpler to measure the number of pedestrians. Data assimilation can be used to calibrate route choice models even without any route choice behavior data. For example, once a route choice model is estimated from the route choice behavior data. Then, at a new event, it may be possible to calibrate the parameters of the route choice model so that the number of people passing through a certain point in the simulation matches the measured data using data assimilation. Since calibrating all the parameters of DCM may drastically change a performance of the model, calibrating only some of the parameters (e.g., the constant term ASC) by data assimilation will be a considerable approach.


\bibliographystyle{abbrv} 
\bibliography{aamas}
\end{document}

%% file: tab/shinkoku_model_parameters_v.tex
\begin{tabular}{lllll|l}
\hline
 $\beta_{\rm DIST}$ & $\beta_{\rm CH}$ & $\beta_{\rm NF}$ & $\beta_{\rm NB}$ & $ASC_{\rm Route2}$ & Accuracy \\ \hline \hline
 -1.33 & 1.13 & 0.202 & -0.105 & 2.33 & 82.2 [\%] \\ \hline
\end{tabular}

%% file: tab/moji_model_parameters.tex
\begin{tabular}{llll|l}
\hline
 $\beta_{\rm DIST}$ & $\beta_{\rm GUIDE}$ & $\beta_{\rm ATT}$ & $ASC_{\rm Route2}$ & Accuracy \\ \hline \hline
 -9.76 & 1.26 & 0.021 & 2.929 & 70.7 [\%] \\ \hline
\end{tabular}

%% file: tab/moji.tex
\begin{tabular}{lllc}
\hline
       & MAE         & RMSE      & Computation time  \\ \hline \hline
Follow & 69.5        & 91.4        & 6 min 21 s \\
DCM    & 54.0 (0.31) & 77.2 (0.33) & 7 min 12 s \\ \hline
\end{tabular}